\documentclass[conference]{IEEEtran}
\IEEEoverridecommandlockouts
\usepackage{cite}
\usepackage{amsmath,amssymb,amsfonts}
\usepackage{epsfig}
\usepackage{algorithmic}
\usepackage{algorithm}
\usepackage{graphicx}
\usepackage{textcomp}
\usepackage{epstopdf}
\usepackage{xcolor}

\def\BibTeX{{\rm B\kern-.05em{\sc i\kern-.025em b}\kern-.08em
    T\kern-.1667em\lower.7ex\hbox{E}\kern-.125emX}}

\begin{document}
\title{Age of Information in Poisson Networks\\
}

\author{\IEEEauthorblockN{Yuming Hu*, Yi Zhong\dag, Wenyi Zhang*}
\IEEEauthorblockA{*Key Laboratory of Wireless-Optical Communications, CAS; University of Science and Technology of China\\
\dag Huazhong University of Science and Technology, China\\
Emails: hymxl@mail.ustc.edu.cn, yzhong@hust.edu.cn, wenyizha@ustc.edu.cn}
}

\maketitle

\begin{abstract}
The age of information (AoI) has been extensively studied in recent years. However, few works have focused on the analysis of AoI in large wireless networks. In this work, we analyze this metric to characterize timeliness in a Poisson bipolar network, in which spatial distribution, fading, and interference are taken into consideration. We further study the effect of deadline constraint on the AoI. We derive upper and lower bounds for the cumulative distribution function (cdf) of the average age in network. Numerical results show that the inclusion of deadline constraint can substantially improve the performance of  the system.

\end{abstract}


\section{Introduction}
\label{sect:introduction}
Numerous applications in networks require the transmission of information about the real-time state between a source and a destination\cite{JKo2010Wireless,Corke2010Environmental}. In these applications, the timeliness of the transmitted message is an important and often critical objective, since an outdated message may be useless. Hence, a theory of \emph{age of information} (AoI) has been recently proposed to characterize the freshness of information at the receiver. The \emph{age} at the time of observation is defined as the current time minus the time at which the observed state (or packet) was generated, and it directly describes the objective of achieving timely updating in a way that traditional metrics (such as delay) do not\cite{Kaul2012Real}.

Many of the previous works analyze the AoI under various system assumptions, with different arrival/departure processes, number of servers, and queue capacities. Server utilizations that minimize age for first-come-first-served (FCFS) \emph{M/M/1, M/D/1, D/M/1} queuing systems were found in \cite{Kaul2012Real}. The average age is shown to decrease as the number of servers increases in \cite{Kam2013Age}. Also, it is shown in \cite{Costa2014Age} that the age with a queue capacity of zero or one can be much lower than an infinite queue capacity, and that replacing packets in the buffer when newer packets arrive performs even better. The AoI in a multi-class \emph{M/G/1} queueing system is considered in \cite{Huang2015Opt}. In \cite{Kam2016Controlling}, the impact of queue sizes ($0$, $1$, and infinity), packet deadlines (fixed and random), and packet replacement on the average age for queuing systems was considered. In our study, we are particularly interested in packet deadline since it reflects the reality that outdated packets may be useless.


The studies above analyze AoI under queuing systems with one queue. There are also some studies that consider AoI for simple networks. The \emph{M/M/1} FCFS system with multiple sources is considered in \cite{Yates2012Real-time}. A wireless broadcast network with a BS sending time-sensitive information to multiple clients over unreliable channels is considered in \cite{Kadota2016Minimizing}. A general multihop network is considered in \cite{Bedewy2017age} and an energy harvesting two-hop network where a source is communicating to a destination through a relay is considered in \cite{Arafa2017Age}.

However, to the best of our knowledge, there has been no work on the AoI in networks considering the locations of transmitters and receivers as well as their interference. The analysis of the average age in spatially random networks is challenging and is further complicated when we consider deadline constraints since the packets may be dropped because of the deadline constraints. In this work, we analyze the AoI in a static Poisson bipolar network, in which spatial distribution, fading, and interference are taken into consideration. We derive upper and lower bounds for the cumulative distribution function (cdf) of the average age in network. We further introduce two kinds of deadline policy and compare them to study the effect of deadline constraints.


\section{System Model}
\label{sect:system model}

We adopt a discrete-time random access system with transmitters and receivers distributed as a Poisson bipolar network. We consider the locations of the transmitters as a Poisson point process (PPP) $\Phi=\{x_i\}\subset\mathbb{R}^2$ of intensity $\lambda$. Each transmitter has a corresponding receiver with a fixed distance $r_0$ but a random orientation. In this work, we consider the typical transmitter located at $x_0\in \Phi$ and $r_0 = |x_0|$ is the distance of this point to the origin, where the corresponding receiver is located. We divide the time into discrete slots with equal duration, and each transmission attempt requires a single time slot. We assume that the locations of all the transmitter-receiver pairs keep unchanged during all time slots once generated at the beginning, i.e., the network is static\cite{zhongyi2016On,zhongyi2017Het}.

The Rayleigh block fading model is considered here and the power fading coefficients remain unchanged within each time slot, and are temporally and spatially independent, with exponential distribution of mean $1$. Denote $\alpha$ as the path loss exponent and $h_{j,x}$ as the power fading coefficient between transmitter $x$ and the corresponding receiver, which is located at the origin $o$, in time slot $j$. We assume that all transmitters transmit at unit power. Focusing on the interference-limited regime with negligible thermal noise, if the signal-to-interference ratio (SIR) over a link is greater than a threshold $\theta$, the packet at the top of the queue of transmitter is successfully transmitted.

We assume that each transmitter maintains an independent queue of a total capacity of two packets (including the packet in transmission). The packets at each transmitter are generated according to a Bernoulli process with arrival rate $\lambda_a$ ($0\leq\lambda_a\leq1$) packets per time slot. Each transmitter attempts to transmit its head-of-line packet with probability $p$ if its queue is not empty in each time slot. The feedback of the status of each attempt of transmission, either successful or failed, is assumed to be instantaneous so that the transmitters are aware of the outcome. If a transmission attempt is successful, the transmitter removes the packet from its queue; otherwise, the transmitter retransmits the packet in the next time slot with probability $p$.


For any time slot $j\in \mathbb{N}^+$, if the typical transmitter is active, the SIR at its corresponding receiver is
\begin{equation}
    \textmd{SIR}_j = \frac{h_{j,x_0}r_0^{-\alpha}}{\sum_{x\in\Phi\backslash\{x_0\}}{h_{j,x}|x|^{-\alpha}\mathbf{1}(x\in\Phi_j)}},
\label{equ:equ_1}
\end{equation}
where $\Phi_j$ is the set of transmitters that are transmitting in that time slot.

Denote the event that the transmission of the typical transmitter $x_0$ succeeds in time slot $j$ conditioned on the PPP $\Phi$ as $\zeta_\Phi^j$. Hence, $\mathbb{P}^{x_0}(\zeta_\Phi^j) = \mathbb{P}( \textmd{SIR}_j >\theta|\Phi,x_0\in\Phi)$ is the success probability of the transmission of the typical transmitter in time slot $j$.

As for the AoI, we define the \emph{age} at the typical receiver as $\Delta_0 (t)=t-g(t)$, where $t$ is the current time slot and $g(t)$ is the time slot at which the most recent successfully transmitted packet, as of time $t$, reached the queue.

\begin{figure}
\centering
\includegraphics[width=0.45\textwidth]{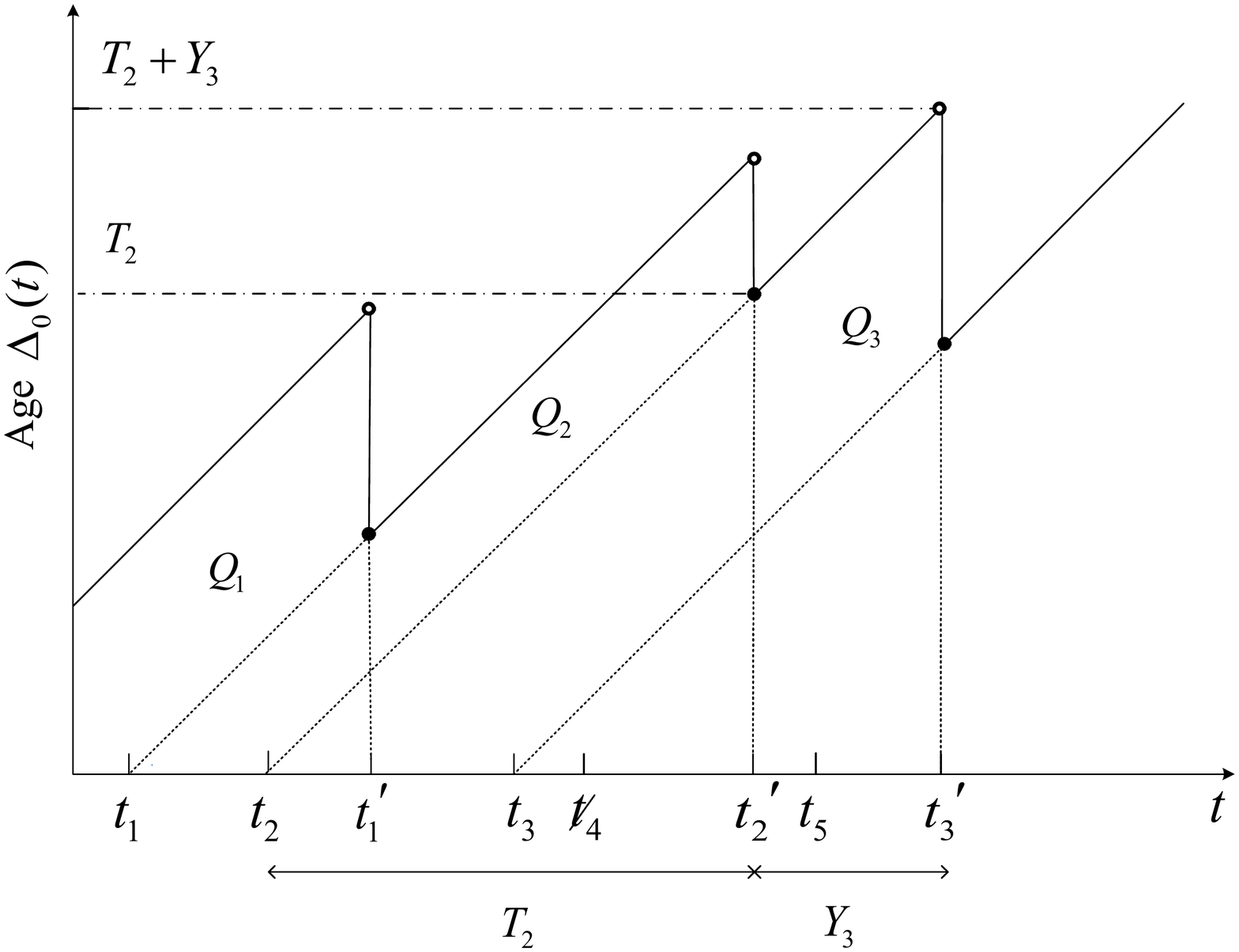}    
\centering
\caption{Age at the typical receiver without deadline.} \label{fig:Age}
\end{figure}

In this work, all the temporal parameters are measured in number of time slots. We assume that the scale of the time is large in Fig. \ref{fig:Age}, \ref{fig:deadlineA} and \ref{fig:deadlineB} to smoothen the curves, so that the properties of the \emph{age} can be observed clearly in these figures.

As shown in Fig. \ref{fig:Age}, packets $1,2,\ldots$ reach the queue at time $t_1,t_2,...$, and are successfully transmitted at time $t_1',t_2',\ldots$. Packet $4$ is dropped because the queue is full when it arrives. $T_k$ is the total time spent in the system from arrival to completing service for the $k$th packet served. We also define the \emph{interdeparture time} $Y_k$ as the time between the instants of completing service for the $(k-1)$th packet served and the $k$th packet served.


Since we adopt a discrete-time system, considering an observation interval $(0,\tau]$. the \emph{time averaged age} is
\begin{equation}
    \Delta_{0/\tau}=\frac{1}{\tau}\sum_{t=1}^{\tau}\Delta_0 (t).
\label{equ:equ_2}
\end{equation}

The sum in \eqref{equ:equ_2} can be divided into sum of the areas of trapezoids $Q_k$ , $k = 1, 2, . . . N(\tau)$. Noting that the area of trapezoid $Q_k$ is $Y_k\cdot [T_{k-1}+(T_{k-1}+Y_k-1)]/2$, we have
\begin{equation}
    \begin{aligned}
        \Delta_{0/\tau}=&\frac{N(\tau)-1}{2\tau} \frac{1}{N(\tau)-1} \sum_{k=2}^{N(\tau)}Y_k\cdot[2T_{k-1}+Y_k-1]\\
                        &+\frac{1}{\tau}\sum_{t=1}^{t_1'-1}\Delta_0 (t).
    \end{aligned}
\end{equation}

The \emph{average age} is defined as we take $\tau$ to infinity
\begin{equation}
    \Delta_0=\lim_{\tau\rightarrow\infty}\Delta_{0/\tau}.
\end{equation}

We define
\begin{equation}
    \lambda_e=\lim_{\tau\rightarrow\infty}{\frac{N(\tau)}{\tau}}
\end{equation}
as the \emph{effective arrival rate}. From \cite{Costa2016On}, due to the ergodicity of $\{Q_k\}$, we have
\begin{equation}
    \begin{aligned}
        \Delta_0=&\lim_{\tau\rightarrow\infty}{ \Delta_{0/\tau}}\\
                =&\lambda_e E\left[Y_k\cdot\frac{1}{2}\cdot(2T_{k-1}+Y_k-1)\right]\\
                =&\lambda_e \left(\frac{1}{2}E[Y_k^2]+E[T_{k-1}Y_k]-\frac{1}{2}E[Y_k]\right).
    \end{aligned}
\label{equ:equ_3}
\end{equation}

We will get the \emph{average age} at the \emph{typical receiver} $\Delta_0$ by deriving $\lambda_e$, $E[Y_k]$, $E[Y_k^2]$ and $E[T_{k-1} Y_k]$, which are related to the success probability $\mathbb{P}^{x_0}(\zeta_\Phi)$.



\section{Age of Information without Deadline}
\label{sect:Age of Information without Deadline}
Before analyzing the \emph{average age}, we need to know the success probability $\mathbb{P}^{x_0}(\zeta_\Phi)$. It is rather difficult to calculate the success probability because of the interacting queueing problem. Therefore, in the following, we obtain upper and lower bounds of the success probability by bringing in two auxiliary systems (a)(b).

In system (a), the typical transmitter behaves exactly the same as in the original system. However, for the other transmitters, when the queue at a transmitter becomes empty, it continues to transmit dummy packets with probability $p$, thus continuing to interfere with other transmissions with probability $p$.

In system (b), the packets will be dropped in interfering transmitters if they are not scheduled by the random access or their transmissions fail.

From \cite{zhongyi2015}, we find that in the auxiliary systems, the success probability for the typical transmitter-receiver pair is the same for each time slot, i.e., the queueing system at the typical transmitter can then be simplified into a Geo/G/1/2 queue (``$2$'' here represents the total capacity of the queue). We denote $\mu$ as the success probability $\mathbb{P}^{x_0}(\zeta_\Phi)$ to simplify the subsequent notation.

Let ${Z_k}$ be the total number of packets in the queue in time slot $k$. Then, $\{Z_k,k\geq0\}$ is a discrete Markov chain, and the range of its values is $\{0,1,2\}$. Its transition probability matrix is
\begin{equation}
    {
        \left[ \begin{array}{ccc}
            1-\lambda_a+\lambda_a\mu & \lambda_a(1-\mu) & 0\\
            (1-\lambda_a)\mu & (1-\lambda_a)(1-\mu)+\lambda_a\mu & \lambda_a(1-\mu)\\
            0 & \mu & 1-\mu
                \end{array}
        \right ]}.
\end{equation}

We get the stationary distribution $\pi=\{\pi_i,i=0,1,2\}$ from the transition probability matrix as follows:
\begin{equation}
\begin{split}
    \pi_0=\frac{(1-\lambda_a)\mu^2}{(1-\lambda_a)\mu^2 + (1-\mu)\lambda_a\mu + (1-\mu)^2\lambda_a^2},   \\
    \pi_1=\frac{(1-\mu)\lambda_a\mu}{(1-\lambda_a)\mu^2 + (1-\mu)\lambda_a\mu + (1-\mu)^2\lambda_a^2},  \\
    \pi_2=\frac{(1-\mu)^2\lambda_a^2}{(1-\lambda_a)\mu^2 + (1-\mu)\lambda_a\mu + (1-\mu)^2\lambda_a^2}.
\end{split}
\label{equ:equ_4}
\end{equation}

Then, we get $\lambda_e$ as
\begin{equation}
    \lambda_e = \lim_{\tau\rightarrow\infty}{\frac{N(\tau)}{\tau}} = \lim_{\tau\rightarrow\infty}{\frac{N(\tau)}{N_0(\tau)}\frac{N_0(\tau)}{\tau}} = (1-P_{\mathrm{los}})\cdot\lambda_a,
\end{equation}
where $N_0(\tau)$ denotes the total number of packets arriving in the queue during $(0,\tau]$, $P_{\mathrm{los}}$ is the percentage of the dropped packets. A packet is dropped if and only if the queue is full when it arrives, so $P_{\mathrm{los}}=\pi_2$, and
\begin{equation}
    \lambda_e = \frac{\lambda_a[(1-\lambda_a)\mu^2+(1-\mu)\lambda_a\mu]}{(1-\lambda_a)\mu^2+(1-\mu)\lambda_a\mu+(1-\mu)^2\lambda_a^2}.
\end{equation}

Before calculating $E[Y_k]$, $E[Y_k^2]$ and $E[T_{k-1} Y_k]$, we denote the event that the $(k-1)$th packet departs the queue leaving behind an empty system as $\psi$ and its complement as $\bar{\psi}$. When a packet is transmitted successfully, the queue may either have one packet or no packet in the queue. So we derive the probability that the event $\psi$ or $\bar{\psi}$ occurs as follows.
\begin{equation}
    \begin{split}
    \textmd{Pr}(\psi) = \frac{\pi_0}{\pi_0+\pi_1} = \frac{(1-\lambda_a)\mu^2}{(1-\lambda_a)\mu^2+(1-\mu)\lambda_a\mu}, \\
    \textmd{Pr}(\bar{\psi}) = 1-\textmd{Pr}(\psi) = \frac{(1-\mu)\lambda_a\mu}{(1-\lambda_a)\mu^2+(1-\mu)\lambda_a\mu}.
\end{split}
\end{equation}

Obviously, when $\psi$ occurs, $E[Y_k|\psi]=\frac{1}{\lambda_a} + \frac{1}{\mu}$, $E[Y_k^2|\psi] = \frac{2(\lambda_a^2+\lambda_a\mu+\mu^2)-\lambda_a\mu(\lambda_a+\mu)}{\lambda_a^2\mu^2}$; when $\bar{\psi}$ occurs, $E[Y_k|\bar{\psi}]=\frac{1}{\mu}$, $E[Y_k^2|\bar{\psi}] = \frac{2-\mu}{\mu^2}$. Thus we get
\begin{equation}
    \begin{split}
    E[Y_k] =& E[Y_k|\psi]\cdot\textmd{Pr}(\psi) +  E[Y_k|\bar{\psi}]\cdot\textmd{Pr}(\bar{\psi})\\
           =&\frac{(\lambda_a+\mu)(1-\lambda_a)\mu+(1-\mu)\lambda_a}{\lambda_a[(1-\lambda_a)\mu^2+(1-\mu)\lambda_a\mu]},
    \end{split}
\end{equation}
\begin{equation}
    \begin{split}
    E[Y_k^2] = E[Y_k^2|\psi]\cdot\textmd{Pr}(\psi) + E[Y_k^2|\bar{\psi}]\cdot\textmd{Pr}(\bar{\psi}).
    \end{split}
\end{equation}

For $T_{k-1}$, it consists of two parts, the waiting time $W_{k-1}$ and the service time $S_{k-1}$. $W_{k-1}$ is the residual service time of the previous packet in service, if there is, so $E[W_{k-1}|\mathrm{ps}]=(1-\mu)\lambda_e/\mu^2$, where ``ps'' means the event that the packet will be served. Besides, we have
\begin{equation}
    \begin{aligned}
    E[S_{k-1}|\psi]&=\sum_{i=1}^{\infty}{i\textmd{Pr}(S_{k-1}=i|\psi)}\\
    &=\sum_{i=1}^{\infty}{\frac{i\textmd{Pr}(S_{k-1}=i,\psi)}{\textmd{Pr}(\psi)}}\\
    &=\frac{1}{\textmd{Pr}(\psi)}\sum_{i=1}^{\infty}{i\textmd{Pr}(S_{k-1}=i,S_{k-1}<X_k)}\\
    &=\frac{(1-\lambda_a)\mu}{[1-(1-\lambda_a)(1-\mu)]^2}\cdot\frac{1}{\textmd{Pr}(\psi)},\\
    \end{aligned}
\end{equation}
\begin{equation}
E[S_{k-1}|\bar{\psi}]=\frac{E[S_{k-1}]-E[S_{k-1}|\psi]\cdot\textmd{Pr}(\psi)}{\textmd{Pr}(\bar{\psi})},
\end{equation}
where $E[S_{k-1}]$ is the average service time which is equal to $1/\mu$.

If the event $\psi$ or $\bar{\psi}$ occurs, $T_{k-1}$ is conditionally independent of $Y_k$, since the event $\psi$ or $\bar{\psi}$ determines whether $Y_k$ is a residual interarrival time plus a service time or just a service time, independent of the time $T_{k-1}$. So we derive $E[T_{k-1}Y_k]$ as below:
\begin{equation}
    \begin{aligned}
    E[T_{k-1}Y_k]=&E[T_{k-1}Y_k|\psi]\textmd{Pr}(\psi)+E[T_{k-1}Y_k|\bar{\psi}]\textmd{Pr}(\bar{\psi})\\
    =&E[T_{k-1}|\psi]E[Y_k|\psi]\textmd{Pr}(\psi)\\  &+E[T_{k-1}|\bar{\psi}]E[Y_k|\bar{\psi}]\textmd{Pr}(\bar{\psi})\\
    =&\left(E[W_{k-1}|\mathrm{ps}]+E[S_{k-1}|\psi]\right)E[Y_k|\psi]\textmd{Pr}(\psi)\\  &+\left(E[W_{k-1}|\mathrm{ps}]+E[S_{k-1}|\bar{\psi}]\right)E[Y_k|\bar{\psi}]\textmd{Pr}(\bar{\psi}).
    \end{aligned}
\label{equ:equ_5}
\end{equation}

Therefore, we get the \emph{average age} of the \emph{typical receiver} $\Delta_0$, which is decided by $\lambda_a$ and $\mu$, i.e., $\mathbb{P}^{x_0}(\zeta_\Phi)$, using the expressions derived in this section.
\begin{equation}
    \Delta_0=\lambda_e \left(\frac{1}{2} E[Y_k^2] + E[T_{k-1} Y_k] - \frac{1}{2}E[Y_k]\right).
\label{equ:equ_6}
\end{equation}

From \cite{zhongyi2015}, we get the cdf of the success probability $\mu=\mathbb{P}^{x_0}(\zeta_\Phi)$ in system (a) as
\begin{equation}
    \begin{aligned}
        \mathbb{P}^{x_0}(\mathbb{P}^{x_0}(\zeta_\Phi)\leq s)=\frac{1}{2}-\frac{1}{\pi}\int_0^\infty \frac{1}{\omega}\textmd{Im}\Bigg\{ p^{j\omega}\exp\bigg(-2\pi\lambda\\
                                                             \int_0^\infty\Big[1-\big(\frac{p}{1+\theta r_0^\alpha r^{-\alpha}}+1-p\big)^{j\omega}\Big]r dr-j\omega\ln s\bigg) \Bigg\}d\omega ,
    \end{aligned}
\label{equ:equ_7}
\end{equation}
which is an upper bound of the cdf of the success probability in the actual system, and also get the cdf of the success probability $\mu=\mathbb{P}^{x_0}(\zeta_\Phi)$ in system (b) as
\begin{equation}
    \begin{aligned}
        \mathbb{P}^{x_0}(\mathbb{P}^{x_0}(\zeta_\Phi)\leq s)=\frac{1}{2}-\frac{1}{\pi}\int_0^\infty \frac{1}{\omega}\textmd{Im}\Bigg\{ p^{j\omega}\exp\bigg(-2\pi\lambda\\
                                                             \int_0^\infty\Big[1-\big(\frac{\lambda_a p}{1+\theta r_0^\alpha r^{-\alpha}}+1-\lambda_a p\big)^{j\omega}\Big]r dr-j\omega\ln s\bigg) \Bigg\}d\omega ,
    \end{aligned}
\label{equ:equ_8}
\end{equation}
which is a lower bound of the cdf of the success probability in the actual system.

By combining (\ref{equ:equ_6}), (\ref{equ:equ_7}) and (\ref{equ:equ_8}), we derive upper and lower bounds of the cdf of the \emph{average age} $\mathbb{P}(\Delta_0\leq t)$.

\section{Age of Information with Deadline}
\label{sect:Age of Information with Deadline}
Different from Section \ref{sect:Age of Information without Deadline}, we introduce two kinds of deadline policy: deadline (A) and (B).

\begin{figure}
\centering
\includegraphics[width=0.45\textwidth]{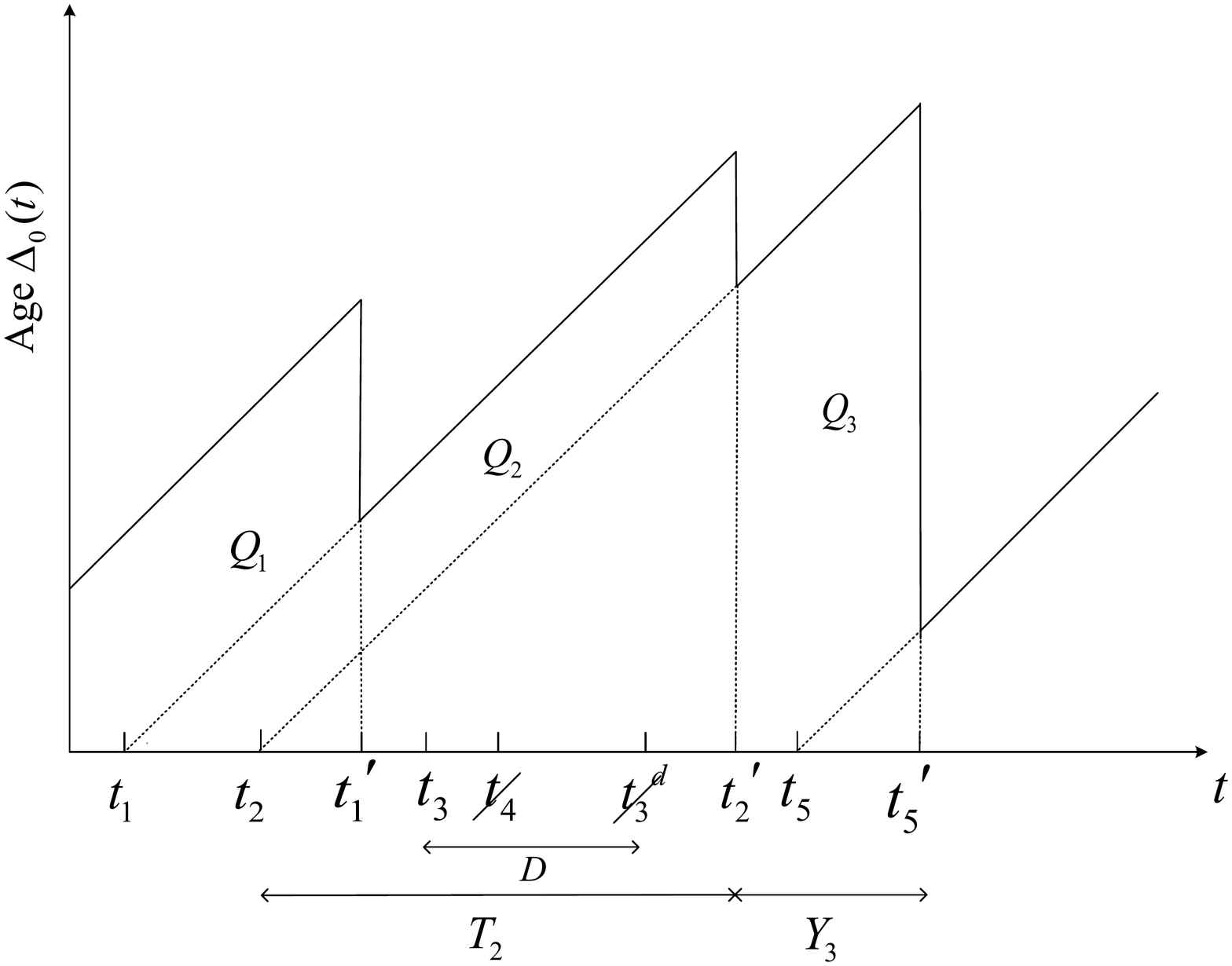}
\centering
\caption{Age at the typical receiver with deadline (A).} \label{fig:deadlineA}
\end{figure}

\begin{figure}
\centering
\includegraphics[width=0.45\textwidth]{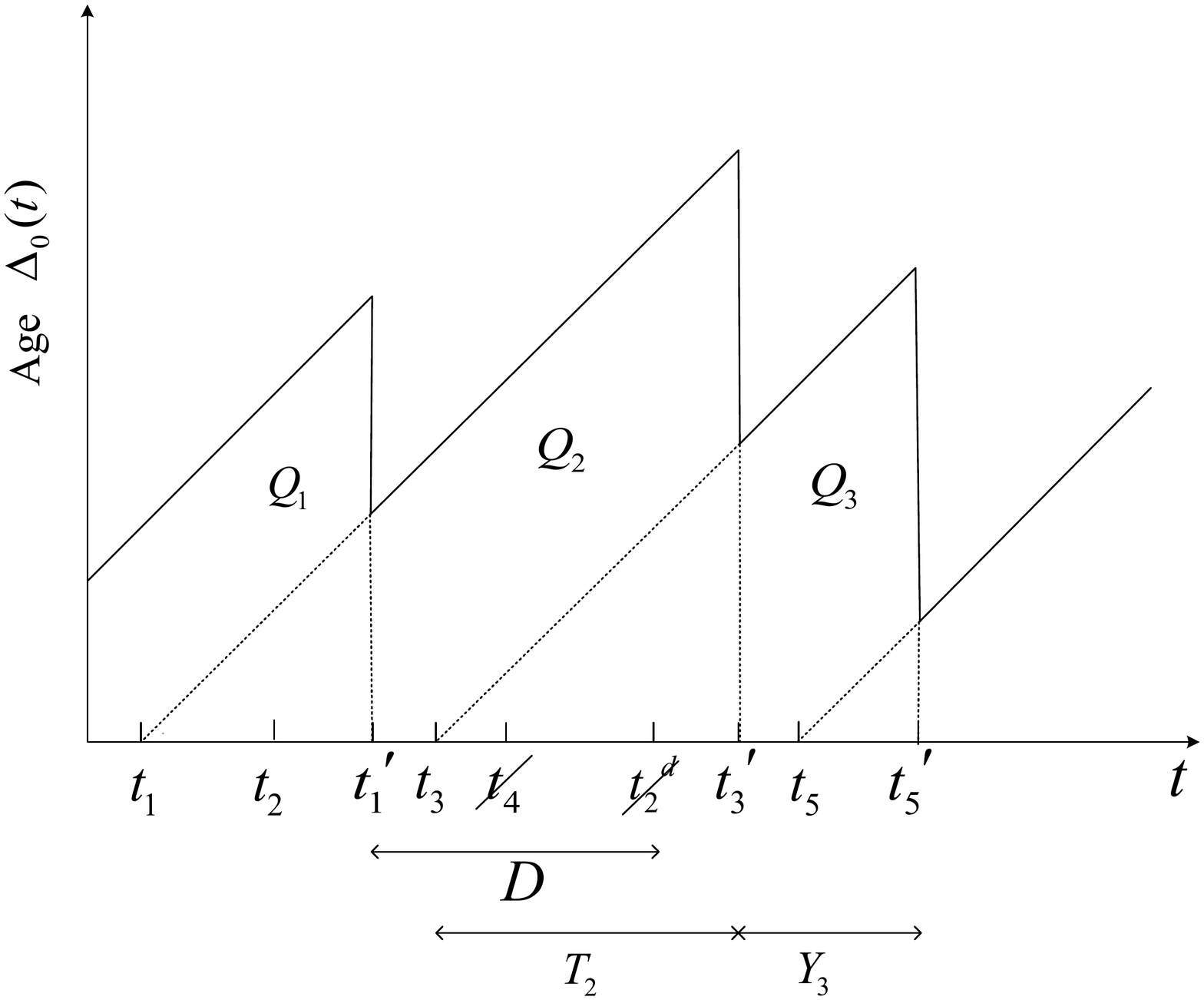}
\centering
\caption{Age at the typical receiver with deadline (B).} \label{fig:deadlineB}
\end{figure}

As shown in Fig. \ref{fig:deadlineA}, with deadline (A), we consider that a packet waiting in the queue is constrained by a deadline, so that if it waits for transmission for a time period longer than $D$, it will be dropped from the system and never enter service, such as packet $3$ in Fig. \ref{fig:deadlineA}. If a packet starts to be transmitted before its deadline expires, it is guaranteed to be served and will never be dropped. As shown in Fig. \ref{fig:deadlineB}, deadline (B) is the case where packets in service can expire, such as packet $2$ in Fig. \ref{fig:deadlineB}, but packets waiting in the queue are never affected. In our work, we let the deadline $D$ be deterministic.

We again need to calculate $\lambda_e$, $E[Y_k]$, $E[Y_k^2]$ and $E[T_{k-1} Y_k]$ to derive the \emph{average age} $\Delta_0$ by (\ref{equ:equ_3}). The analysis is similar to that in Section \ref{sect:Age of Information without Deadline}, so only the differences will be presented in this section. \

The total number of packets in the queue is no longer a discrete Markov chain when we add deadline to the model. Following \cite{Kam2016Age}, we adopt a time averaging approach to calculate $\Delta_0$. The queue of packets has three states $\{0,1,2\}$, respectively representing there is (are) $0,1$ or $2$ packet(s) in the queue. Denoting by $E[V_i]$ the average time spent in state $i$ each time the queue entering that state, and by $p_i$ the percentage of time spent in state $i$ during the whole time.

From \cite{Kam2016Age}, we get
\begin{equation}
\begin{split}
    p_0=\frac{\pi_0 E[V_0]}{\pi_0 E[V_0]+\pi_1 E[V_1]+\pi_2 E[V_2]},   \\
    p_1=\frac{\pi_1 E[V_1]}{\pi_0 E[V_0]+\pi_1 E[V_1]+\pi_2 E[V_2]},  \\
    p_2=\frac{\pi_2 E[V_2]}{\pi_0 E[V_0]+\pi_1 E[V_1]+\pi_2 E[V_2]},
\end{split}
\label{equ:equ_9}
\end{equation}
where $\{\pi_i\}$ are given in (\ref{equ:equ_4}).

\subsection{Deadline (A)}
\label{subsect:Deadline A}
With deadline (A), $V_0$ is simply an interarrival time, i.e., $E[V_0]=1/\lambda_a$. As for $E[V_1]$ and $E[V_2]$, they are given by
\begin{equation}
    \begin{aligned}
        E[V_1]=&E[t_S|t_S \leq t_A]\textmd{Pr}(t_S \leq t_A) \\
               &+ E[t_A|t_A < t_S]\textmd{Pr}(t_A < t_S)\\
              =&\frac{\mu+(1-\mu)\lambda_a}{[1-(1-\mu)(1-\lambda_a)]^2},
    \end{aligned}
\label{equ:equ_10}
\end{equation}
\begin{equation}
    \begin{aligned}
        E[V_2]=&E[t_S|t_S \leq D]\textmd{Pr}(t_S \leq D)\\
               &+ E[D|D < t_S]\textmd{Pr}(D < t_S)\\
              =&\frac{1+(\mu^2 D+\mu-1)(1-\mu)^D}{\mu},
    \end{aligned}
\label{equ:equ_11}
\end{equation}
where $t_S$ is the residual service time of the packet transmitted currently, and $t_A$ is the residual arrival time of the packet which will arrive in the queue next.

$E[Y_k|\psi]$, $E[Y_k|\bar{\psi}]$, $E[Y_k^2|\psi]$ and $E[Y_k^2|\bar{\psi}]$ are the same as those in Section \ref{sect:Age of Information without Deadline}, and
\begin{equation}
\begin{split}
    E[Y_k] = E[Y_k|\psi]\cdot\textmd{Pr}(\psi) +  E[Y_k|\bar{\psi}]\cdot\textmd{Pr}(\bar{\psi}),\\
    E[Y_k^2] = E[Y_k^2|\psi]\cdot\textmd{Pr}(\psi) + E[Y_k^2|\bar{\psi}]\cdot\textmd{Pr}(\bar{\psi}),
\end{split}
\label{equ:equ 12}
\end{equation}
where $\textmd{Pr}(\psi)$ and $\textmd{Pr}(\bar{\psi})$ are given by
\begin{equation}
    \begin{aligned}
        \textmd{Pr}(\psi) &= \frac{p_0}{p_0+p_1},\\
        \textmd{Pr}(\bar{\psi}) &= 1-\textmd{Pr}(\psi).
    \end{aligned}
\label{equ:equ_13}
\end{equation}

Since packets in the queue are constrained by deadline (A), a packet is dropped if the queue is full when it arrives or if it comes into the queue while there is one packet under transmission, but still waits for transmission for a time period longer than $D$. So
\begin{equation}
        P_{\mathrm{los}} =p_2+p_1(1-\mu)^D.
\end{equation}

Then, we have
\begin{equation}
        \lambda_e =(1-P_{\mathrm{los}})\cdot\lambda_a = \lambda_a[p_0+p_1(1-(1-\mu)^D)].
\end{equation}

The average waiting time is $E[W_{k-1}|\mathrm{ps}]=(1-\mu)\lambda_e/\mu^2$, and
\begin{equation}
    \begin{aligned}
    E[S_{k-1}|\psi]=&\sum_{i=1}^{\infty}\sum_{j=0}^{i}\frac{{i\textmd{Pr}(S_{k-1}=i, j\mbox{ packets dropped})}}{\textmd{Pr}(\psi)}\\
    =&\frac{(1-\lambda_a)\mu}{\textmd{Pr}(\psi)(1-\lambda_a-\lambda_a P_{\mathrm{los}})}\Big[\frac{1-\lambda_a}{1-(1-\lambda_a)(1-\mu)}\\
     &-\frac{\lambda_a^2 P_{\mathrm{los}}^2}{(1-\lambda_a)(1-(1-\mu)\lambda_a P_{\mathrm{los}})}\Big],\\
    \end{aligned}
\end{equation}
\begin{equation}
E[S_{k-1}|\bar{\psi}]=\frac{E[S_{k-1}]-E[S_{k-1}|\psi]\cdot\textmd{Pr}(\psi)}{\textmd{Pr}(\bar{\psi})},
\end{equation}
where $E[S_{k-1}]$ is the average service time which is equal to $1/\mu$.

Since (\ref{equ:equ_5}) and (\ref{equ:equ_6}) still hold, we can get $E[T_{k-1}Y_k]$ and then derive $\Delta_0$ under deadline (A).

\subsection{Deadline (B)}
With deadline (B), $V_0$ is also simply an interarrival time, i.e., $E[V_0]=1/\lambda_a$, and
\begin{equation}
    \begin{aligned}
        E[V_1]=&E[t_S|t_S \leq t_A,t_S \leq D]\textmd{Pr}(t_S \leq t_A,t_S \leq D)\\
               &+E[t_A|t_A < t_S,t_A \leq D]\textmd{Pr}(t_A < t_S,t_A \leq D)\\
               &+E[D|D < t_S,D < t_A]\textmd{Pr}(D < t_S,D < t_A)\\
              =&\frac{[\lambda_a(1-\mu)+\mu][1-(D+1)(1-\mu)^D(1-\lambda_a)^D]}{[1-(1-\mu)(1-\lambda_a)]^2}\\
               &+\frac{[\lambda_a(1-\mu)+\mu][D(1-\mu)^{D+1}(1-\lambda_a)^{D+1}]}{[1-(1-\mu)(1-\lambda_a)]^2}\\
               &+D(1-\mu)^D(1-\lambda_a)^D,
    \end{aligned}
\end{equation}
\begin{equation}
    \begin{aligned}
        E[V_2]=&E[t_S|t_S \leq D]\textmd{Pr}(t_S \leq D)\\
               &+ E[D|D < t_S]\textmd{Pr}(D < t_S)\\
              =&\frac{1+(\mu^2 D+\mu-1)(1-\mu)^D}{\mu}.
    \end{aligned}
\end{equation}

$E[Y_k|\psi]$, $E[Y_k|\bar{\psi}]$, $E[Y_k^2|\psi]$ and $E[Y_k^2|\bar{\psi}]$ here are the same as those in Section \ref{sect:Age of Information without Deadline}, and (\ref{equ:equ 12}), (\ref{equ:equ_13}) still hold here.

Since packets in the queue are constrained by deadline (B), a packet is dropped if the queue is full when it arrives or it comes into the queue while there is one or no packet under transmission, but its transmission time is longer than $D$. Thus
\begin{equation}
        P_{\mathrm{los}} =p_2+(p_0+p_1)(1-\mu)^D.
\end{equation}

Then we derive
\begin{equation}
        \lambda_e =(1-P_{\mathrm{los}})\cdot\lambda_a = \lambda_a(p_0+p_1)[1-(1-\mu)^D].
\end{equation}

The average waiting time is
\begin{equation}
    \begin{aligned}
        E[W_{k-1}|\mathrm{ps}]=&E[W_{k-1}|S_{k-2} < D,\mathrm{ps}]\textmd{Pr}(S_{k-2} < D)\\
                       &+ E[W_{k-1}|S_{k-2} \geq D,\mathrm{ps}]\textmd{Pr}(S_{k-2} \geq D)\\
                      =&\frac{1-\mu}{\mu^2}\cdot(1-(1-\mu)^{D-1})\lambda_e\\
                       &+ \frac{D(D-1)}{2}\cdot(1-\mu)^{D-1}\lambda_e.
    \end{aligned}
\end{equation}

Also we have
\begin{equation}
    \begin{aligned}
    E[S_{k-1}|\psi]=&\frac{1}{\textmd{Pr}(\psi)}\sum_{i=1}^{D}{i\textmd{Pr}(S_{k-1}=i\leq D,S_{k-1}<X_k)}\\
    =&\bigg\{\frac{\mu(1-\lambda_a)[1-(D+1)(1-\mu)^D(1-\lambda_a)^D]}{[1-(1-\mu)(1-\lambda_a)]^2}\\
     &+ \frac{D\mu(1-\lambda_a)(1-\mu)^{D+1}(1-\lambda_a)^{D+1}}{[1-(1-\mu)(1-\lambda_a)]^2}\bigg\}\cdot\frac{1}{\textmd{Pr}},
    \end{aligned}
\end{equation}
\begin{equation}
E[S_{k-1}|\bar{\psi}]=\frac{E[S_{k-1}]-E[S_{k-1}|\psi]\cdot\textmd{Pr}(\psi)}{\textmd{Pr}(\bar{\psi})},
\end{equation}
where $E[S_{k-1}]$ is the average service time which is equal to $1/\mu$.

Since (\ref{equ:equ_5}) and (\ref{equ:equ_6}) still hold here, we can get $E[T_{k-1}Y_k]$ and then derive $\Delta_0$ under deadline (B).

\section{Numerical Results}
\label{sect:Simulation Results}
\begin{figure}
\centering
\includegraphics[width=0.48\textwidth]{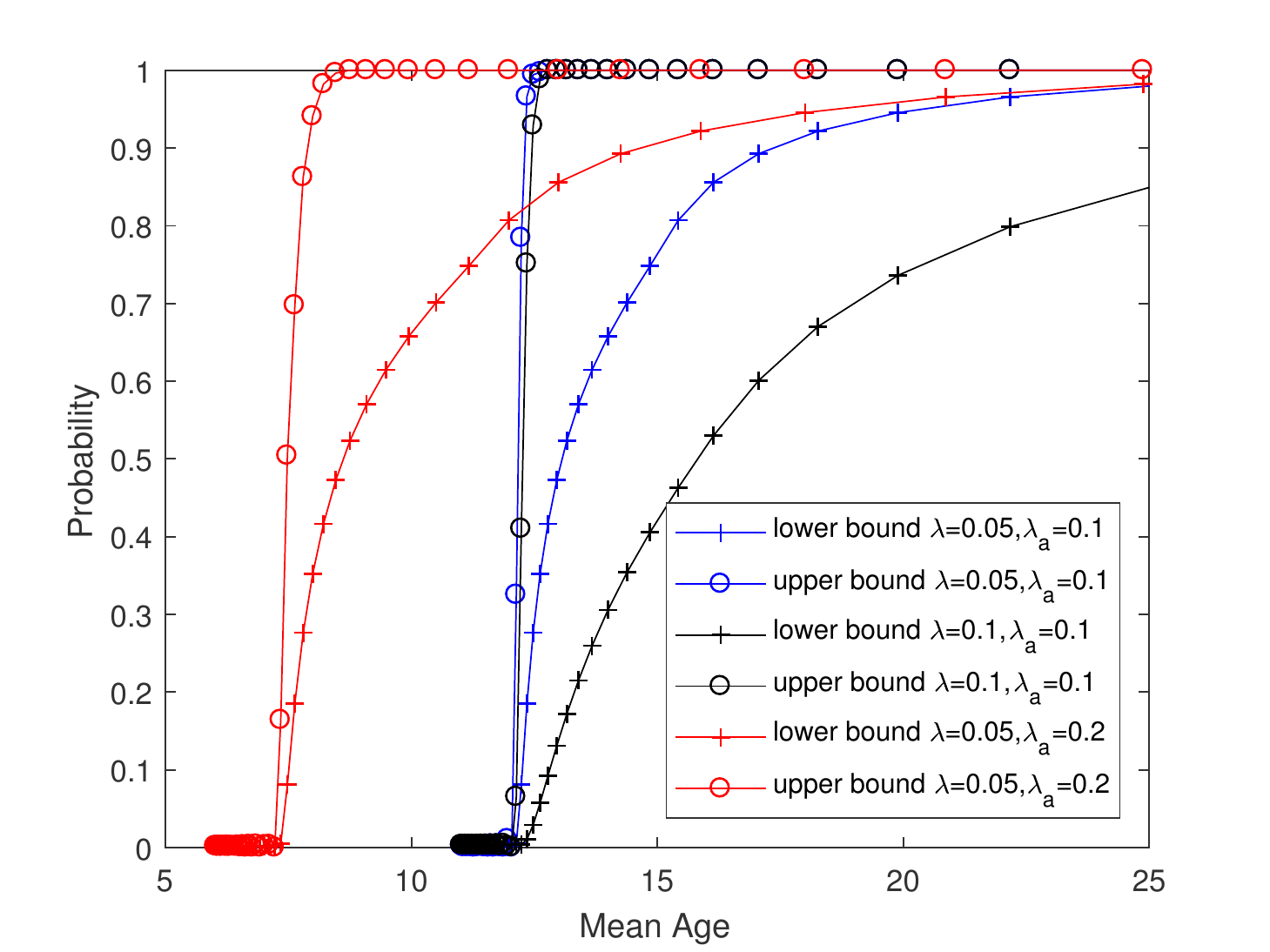}    
\centering
\caption{Lower and upper bounds for the cdf of mean age without deadline for different transmitter intensities $\lambda$ and packet arrival rates $\lambda_a$. The parameters are set as $\theta=10\textmd{dB}$, $r_0=1$, $p=0.5$ and $\alpha=4$.} \label{fig:age_nodeadline}
\end{figure}

\begin{figure}
\centering
\includegraphics[width=0.48\textwidth]{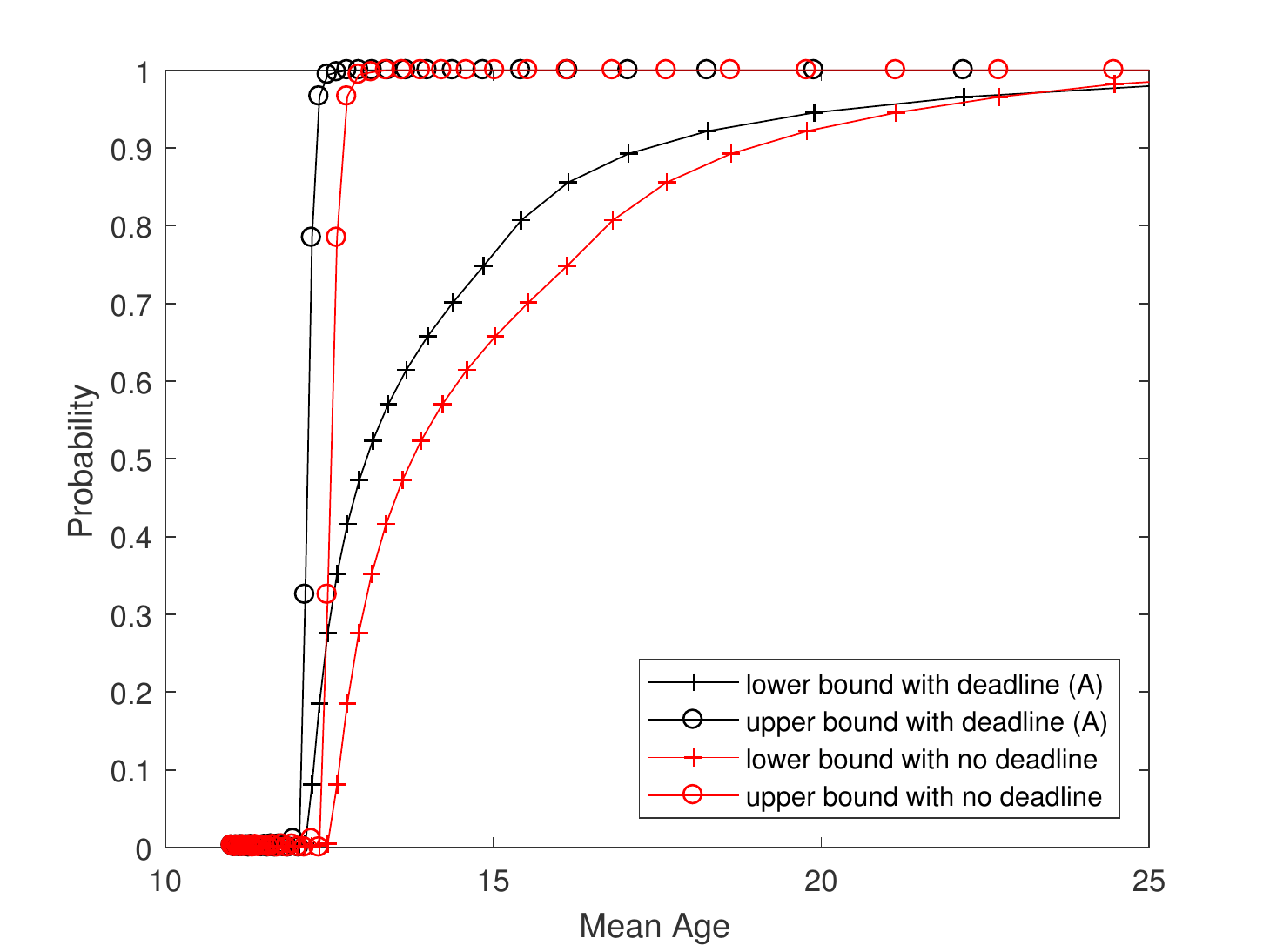}    
\centering
\caption{Comparison of the cdf of mean age with deadline (A) and that without deadline. $\lambda=0.05$, $\lambda_a=0.1$ and the deadline is set as $D=10$.} \label{fig:deadlineA_vs_nodead}
\end{figure}

\begin{figure}
\centering
\includegraphics[width=0.48\textwidth]{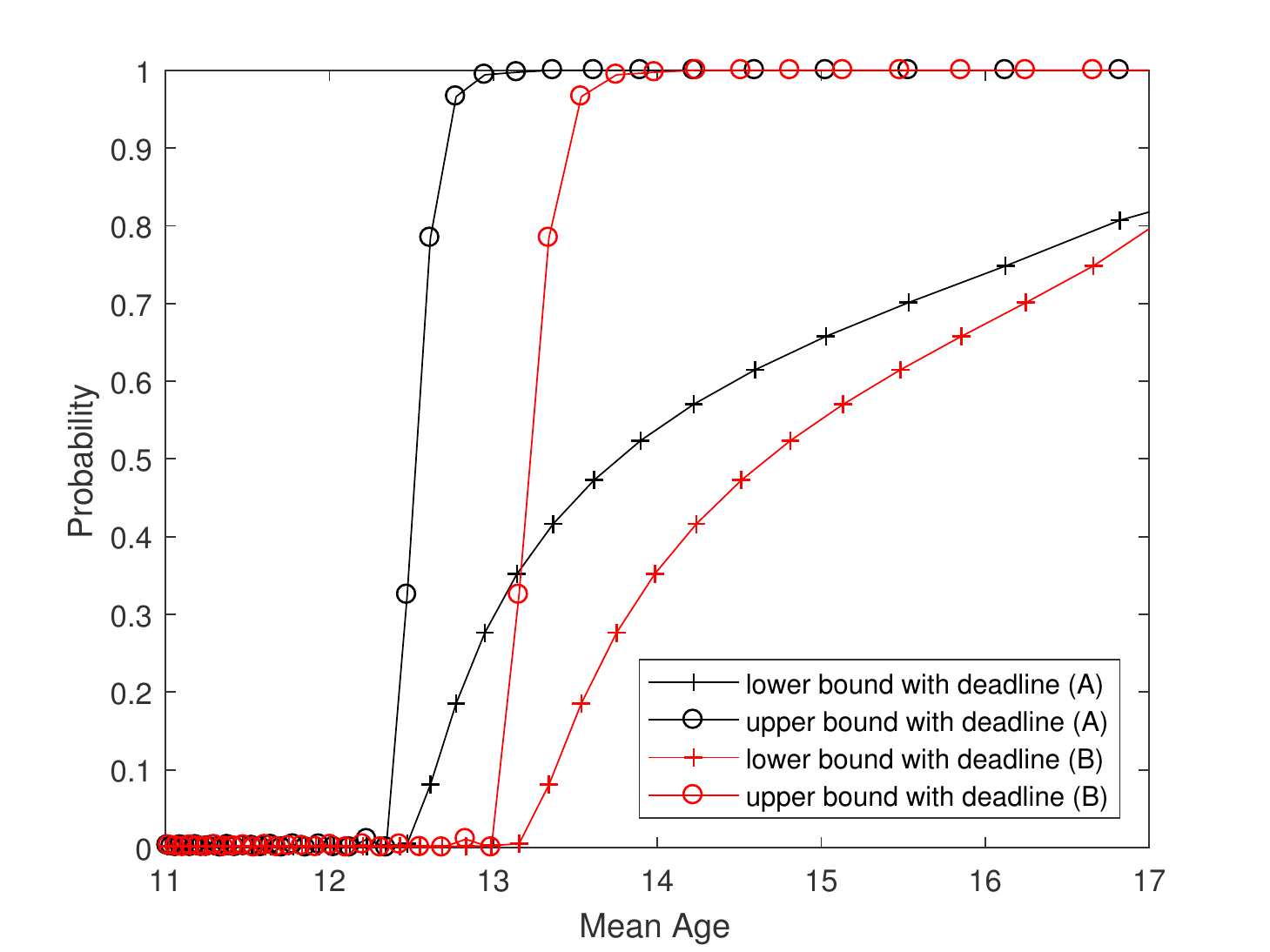}    
\centering
\caption{Comparison of the cdf of mean age with deadline (A) and with deadline (B). $\lambda=0.05$, $\lambda_a=0.1$ and the deadline is set as $D=10$.} \label{fig:deadlineA_vs_B}
\end{figure}

Fig. \ref{fig:age_nodeadline} plots the lower and upper bounds for the cdf of mean age without deadline for different transmitter intensities $\lambda$ and packet arrival rates $\lambda_a$. It shows that the lower and upper bounds are close when $\lambda=0.05$ and $\lambda_a=0.1$, but become loose if $\lambda$ or $\lambda_a$ increases since the difference between the auxiliary systems increases.

Moreover, we observe that the mean age increases when $\lambda$ increases due to the increased interference, and that the mean age decreases when $\lambda_a$ increases, because more packets arrive and the packets  are received more frequently.

Fig. \ref{fig:deadlineA_vs_nodead} plots the lower and upper bounds for the cdf of mean age with deadline (A) and without deadline. It shows that the use of an appropriate packet deadline (A), which constrains the waiting times in the queue, can make the mean age decrease.

Fig. \ref{fig:deadlineA_vs_B} plots the lower and upper bounds for the cdf of mean age with deadline (A) and with deadline (B). It shows that the mean age with deadline (A) is smaller than that with deadline (B) under the used parameters. The gap between the mean ages with deadlines (A) and (B) appears because there is a chance that the transmission time of a packet is longer than $D$, causing the packet to be dropped due to deadline (B), but there is no packet waiting in the queue and then the transmitter has no packet to transmit for a while. This occurs with a low but non-negligible probability with deadline (B) but never occurs with deadline (A).

\section{Conclusion}
In this paper, we investigated the AoI of a static Poisson bipolar network. We derive upper and lower bounds for the cdf of the mean age without deadline, with deadline (A) and deadline (B), respectively. Our results show that the use of an appropriate packet deadline policy can reduce the mean age in network, and deadline (A), which affects the waiting times in the queue, is usually better than deadline (B), which constrains the packets under transmission.

\section*{Acknowledgment}
This work was supported in part by the Key Research Program of Frontier Sciences of CAS under Grant QYZDYSSW-JSC003, by the National Natural Science Foundation of China under Grants 61722114 and 61701183, and by the Fundamental Research Funds for the Central Universities under Grants WK3500000003, WK3500000005 and 2018KFYYXJJ139.

\end{document}